\documentclass[aps,prb,twocolumn]{revtex4-1}

\usepackage{graphicx}
\usepackage{dcolumn}
\usepackage{bm}
\usepackage{braket}
\usepackage{dsfont}
\usepackage{bbold}
\usepackage{amsmath}
\usepackage{subfigure}
\usepackage{extarrows}
\usepackage{amssymb}
\usepackage{esint}
\usepackage{qcircuit}
\usepackage{tikz}
\usepackage{hyperref}
\usepackage{xcolor}
\usepackage{lipsum}
\usetikzlibrary{trees,arrows}

\begin{document}

\newcommand{\redmark}[1] {\color{red}\textbf{#1}\color{black}\normalsize}
\newcommand{\bluemark}[1] {\color{blue}\textbf{#1}\color{black}\normalsize}
\newcommand{\brownmark}[1] {\color{purple}\textbf{#1}\color{black}\normalsize}
\newcommand{\uvec}[1]{\boldsymbol{\hat{\textbf{#1}}}}

\title{Gutzwiller wave function on a digital quantum computer}

\author{Bruno Murta}
\email{bpmurta@gmail.com}
\affiliation{QuantaLab, International Iberian Nanotechnology Laboratory (INL), 4715-330 Braga, Portugal}
\affiliation{Departamento de F\'{i}sica, Universidade do Minho, 4710-057 Braga, Portugal}
\author{J. Fern\'{a}ndez-Rossier}
\affiliation{QuantaLab, International Iberian Nanotechnology Laboratory (INL), 4715-330 Braga, Portugal}
\affiliation{Departamento de F\'isica Aplicada, Universidad de  Alicante, San Vicente del Raspeig 03690, Spain}

\date{\today}

\begin{abstract}
The determination of the ground state of quantum many-body systems via digital quantum computers rests upon the initialization of a sufficiently educated guess. This requirement becomes more stringent the greater the system. Preparing physically-motivated ans\"{a}tze on quantum hardware is therefore important to achieve quantum advantage in the simulation of correlated electrons. In this spirit, we introduce the Gutzwiller Wave Function (GWF) within the context of the digital quantum simulation of the Fermi-Hubbard model. We present a quantum routine to initialize the GWF that comprises two parts. In the first, the noninteracting state associated with the $U=0$ limit of the model is prepared. In the second, the non-unitary Gutzwiller projection that selectively removes states with doubly-occupied sites from the wave function is performed by adding to every lattice site an ancilla qubit, the measurement of which in the $|0\rangle$ state confirms the projection was made. Due to its non-deterministic nature, we estimate the success rate of the algorithm in generating the GWF as a function of the lattice size and the interaction strength $U/t$. The scaling of the quantum circuit metrics and its integration in general quantum simulation algorithms are also discussed.
\end{abstract}

\pacs{Valid PACS appear here}

\maketitle

The quantum many-body problem permeates a wide range of fields of research within condensed matter physics, quantum chemistry, and materials science. In particular, it is the cornerstone of the electronic structure problem. Conventional-hardware-based numerical methods have played a pivotal role in unravelling the electronic structure of materials, but not without shortcomings. Indeed, although the low-energy properties of weakly interacting materials are well described by Density-Functional Theory\cite{Hohenberg_Kohn_64, Kohm_Sham_65} with approximate functionals based on the Local Density Approximation\cite{Lewin_Lieb_Seiringer19}, this approach often fails when strong electron-electron interactions prevail. Quantum Monte Carlo\cite{Foulkes01} methods are a leading alternative, though often plagued by the sign problem\cite{Troyer_Wiese05}.

In principle, a more accurate description of correlated fermions could be achieved via wave-function-based methods, but the storage and manipulation of the wave function in classical hardware is hampered by the exponential wall problem\cite{Kohn99}. 
This is, however, not the case in quantum hardware, thanks to the principle of superposition and the natural encoding of entanglement. Quantum computers have thus been proposed\cite{Feynman82} as a platform to simulate quantum many-body models that encapsulate the electronic structure of materials when their understanding demands an explicit representation of the wave function, either because of the presence of strong correlations or a high accuracy requirement\cite{Cao19}.

A number of quantum algorithms to determine the ground state of a given Hamiltonian $\mathcal{H}$ have been put forth. The most prominent example is Quantum Phase Estimation\cite{Kitaev95} (QPE), whereby an initial state with non-negligible overlap with the exact ground state undergoes time evolution under the action of the propagator $e^{-i\mathcal{H}t/\hbar}$ subject to the control of ancilla qubits, from which the eigenspectrum can be extracted after the application of the inverse quantum Fourier transform. In particular, once the ground state energy is read out from the ancilla qubits, the state collapses into the exact ground state. 

Because the resources required to implement QPE are far beyond the capacity of near-term quantum processors\cite{cruz20}, a leaner class of hybrid variational algorithms\cite{Cerezo20} have been developed, of which the Variational Quantum Eigensolver\cite{Peruzzo14} (VQE) is the reference. In VQE, a parameterized state is prepared on a quantum computer, which is used to compute the expectation value of $\mathcal{H}$. This energy is then provided to a classical computer that performs the optimization routine to find the updated parameter values, which are then fed back to the quantum computer to begin the next iteration. Alternatively, the ground state can be found via Quantum Imaginary Time Evolution (QITE), which can be implemented in quantum hardware by casting it into a variational problem\cite{McArdle191} or by finding the unitary operation\cite{Motta20} that transforms the state at the current step, $\ket{\psi(\tau)}$, into the (normalized) state at the next step, $\ket{\psi(\tau + \Delta \tau)} = \mathcal{N} e^{-\mathcal{H} \Delta \tau/ \hbar} \ket{\psi(\tau)}$. 

Despite the development of the aforementioned quantum simulation algorithms, it is well established\cite{Kempe04} that the problem of finding the ground state of many Hamiltonians involving only local interactions is QMA-complete\footnote{QMA, short for \textit{Quantum Merlin-Arthur}, is the quantum analog of the nonprobabilistic complexity class NP.}. Regarding QPE and QITE, the key challenge lies in the preparation of the initial state: the scaling of both algorithms is polynomial with respect to the inverse of the overlap between the initial state and the exact ground state\cite{Ge_Tura_Cirac19}, so the state preparation routine should lead to a polynomially decreasing overlap as the system size increases for the overall algorithm to be efficient\cite{McArdle20}. Nevertheless, conventional choices of initial states, such as noninteracting or mean-field ground states, produce an exponentially vanishing probability of collapsing into the ground state due to the orthogonality catastrophe\cite{Anderson67}. This overlap can in principle be enhanced via adiabatic evolution\cite{Aspuru-Guzik05}, but its success depends on the gap between the ground and first excited states throughout the adiabatic path, which is generally unknown. 

VQE, in turn, is a heuristic, so, contrary to QPE and QITE, there is no theoretical guarantee of its success given an appropriate initial state. In fact, the challenge in VQE is not quite preparing the initial state, but rather formulating a parameterized ansatz such that its manifold includes a path connecting the initial state to the exact ground state without incurring in an exponential number of parameters. In any case, preparing an initial state with a greater overlap with the exact ground state could simplify the variational procedure by shortening the path in parameter space that needs to be covered, potentially reducing the number of optimization steps or the layers of the ansatz. Perhaps even more important is the possible avoidance of barren plateaus\cite{McClean18}, which can be linked to an uninformed initialization of the ansatz\cite{Holmes21}.

Given the crucial role played by the initial state in the success of digital quantum simulation schemes, developing routines to prepare on quantum hardware physically-motivated ans\"{a}tze is of great importance. It is within this context that, in this paper, we propose an algorithm to prepare on quantum hardware the Gutzwiller Wave Function\cite{Gutzwiller63} (GWF). This is a conceptually simple, physically-motivated ansatz that contains a single parameter. It can be applied to the wide class of lattice models of correlated electrons where the electron-electron Coulomb repulsion is described, to leading order, in terms of the on-site Hubbard interaction\cite{Hubbard63} $U \hat{n}_{i,\uparrow} \hat{n}_{i,\downarrow}$, where $\hat{n}_{i,\alpha} \equiv \hat{a}^{\dagger}_{i,\alpha} \hat{a}_{i,\alpha}$ is the number operator acting on site $i$ for electrons of spin $\alpha = \; \uparrow, \downarrow$, with $\hat{a}_{i,\alpha}$ the annihilation operator in second quantization.

\begin{figure*}
\includegraphics[width=0.9\linewidth]{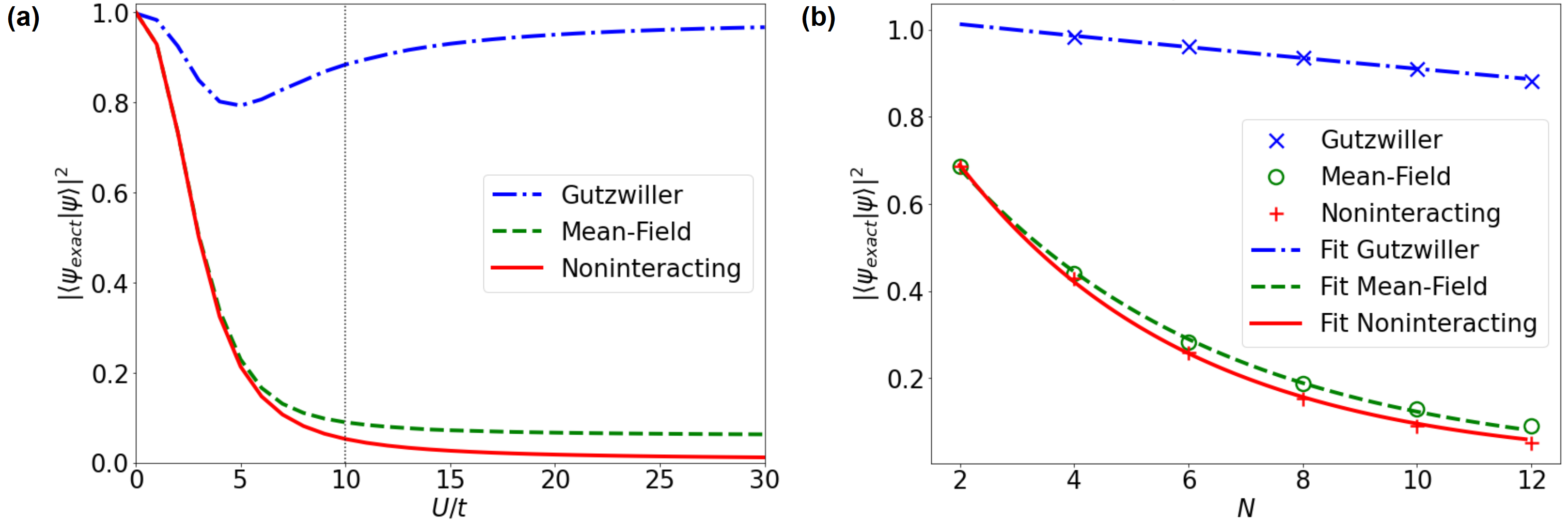}
\caption{Comparison of Gutzwiller Wave Function (GWF) to noninteracting and self-consistent mean-field ground states for Fermi-Hubbard Model (FHM) in chain with open boundary conditions at half-filling. (a) Fidelity of three reference states with respect to exact ground state for a chain of $N = 12$ sites against normalized Hubbard parameter $U/t$. (b) Same as (a), but now size of chain $N$ is varied between 2 and 12 sites, while $U/t$ takes the fixed value of $10$, as highlighted in (a) by the vertical dotted line. Numerical results were fitted to exponential decay $c_1 e^{-c_2 N}$.
\label{fig1}}
\end{figure*}

The Gutzwiller Wave Function (GWF) is defined as
\begin{equation}
    \ket{\psi_G} = \prod_{i = 1}^{N} (\mathbb{1} - g \hat{n}_{i,\uparrow} \hat{n}_{i,\downarrow}) \ket{\psi_0},
\end{equation}
\noindent where $g \in [0,1]$ is a free parameter, $N$ is the number of lattice sites, and $\ket{\psi_0}$ is the noninteracting ground state. In words, the GWF is prepared by reducing the amplitude of the basis states of $\ket{\psi_0}$ with doubly-occupied sites. The degree by which the amplitude is decreased is set by $g$. The optimal value of $g$ for a given Hubbard parameter $U$ is found by minimizing the energy. The greater the magnitude of the on-site Hubbard interactions, the more unfavorable the doubly-occupied states are, and hence the greater $g$ is. In particular, when $U = 0$, $g = 0$, and when $U \to \infty$, $g = 1$. Importantly, the Gutzwiller projection does not break any symmetry of the Hamiltonian.  

In spite of its apparent simplicity, the GWF is a complex state that captures some correlations between the electrons. This complexity can be understood as follows. The noninteracting ground state 
\begin{equation}
\ket{\psi_0}=\prod_{\alpha,\sigma} a^{\dagger}_{\alpha,\sigma}\ket{0} 
\end{equation}
is a single Slater determinant, or a Fock state, when the single-particle basis $\{ |\alpha\rangle \}$ is chosen to be the eigenbasis of the noninteracting tight-binding Hamiltonian. However, the Gutzwiller projection is carried out in the site basis, therefore the operators $\{ a^{\dagger}_{\alpha, \sigma}\}$ have to be expanded as a linear combination of site operators $\{ a^{\dagger}_{i, \sigma}\}$,
\begin{equation}
\ket{\psi_0}=\prod_{\alpha,\sigma} \left(\sum_{i} \phi_\alpha(i) a^{\dagger}_{i, \sigma}\right)\ket{0},
\end{equation}
in which case $\ket{\psi_0}$ appears as multi-determinant state. 

As a result, the determination of expectation values of the GWF is a many-body problem that cannot be solved exactly except in the special cases of one\cite{metzner87,gebhard87} and infinite dimensions\cite{Gutzwiller65}. Hence, in two and three dimensions, numerical methods such as Variational Monte Carlo\cite{Gros87, Yokoyama87} have been employed to compute expectation values of the GWF. The GWF has been used to model a variety of correlated fermion problems, such as the metal-insulator transition\cite{brinkman70}, the low-temperature behaviour of  $^3\textrm{He}$\cite{vollhardt84}, and the superconductivity in the cuprates\cite{Anderson87}.

In order to appreciate the significance of the Gutzwiller projection as an improved starting point for general quantum simulation algorithms, we  consider  the Fermi-Hubbard Model (FHM), 
\begin{equation}
    \hat{\mathcal{H}} = -t \sum_{i = 1}^{N}\sum_{\sigma = \uparrow, \downarrow} (\hat{a}^{\dagger}_{i,\sigma} \hat{a}_{i+1,\sigma} + \textrm{H.c.}) + U \sum_{i = 1}^N \hat{n}_{i, \uparrow} \hat{n}_{i, \downarrow},
\end{equation}
at half-filling in a one-dimensional lattice with open boundary conditions and with up to $N=12$ sites. Specifically, we compare the fidelity with respect to the exact ground state of three trial states: the GWF, the noninteracting and the self-consistent mean-field ground states. The exact ground state is obtained via exact diagonalization using the numerical package QuSpin \cite{QuSpin}. The self-consistent mean-field theory amounts to a direct Hartree-Fock decoupling of the quartic Hubbard term, which gives rise to coefficients that depend on the expectation values of occupation numbers. Self-consistency is attained when single-particle occupations of the mean-field ground state coincide (to a given precision) with the input values. Random initial conditions are used, and the lowest-energy state out of all trials is selected. 

Fig. 1(a) presents the fidelity of these three states with respect to the exact ground state for a chain of $N = 12$ sites; similar profiles are observed for chains of different sizes. For concreteness, Fig. 1(b) shows how the fidelity of the three states decays with the size of the chain for the specific value of $U/t = 10$. The decrease of the overlap between the GWF and the exact ground state is significantly slower, thus rendering it a far better starting point for QPE, VQE or QITE than the single-particle states, especially for large systems.

\begin{figure*}
\includegraphics[width=\linewidth]{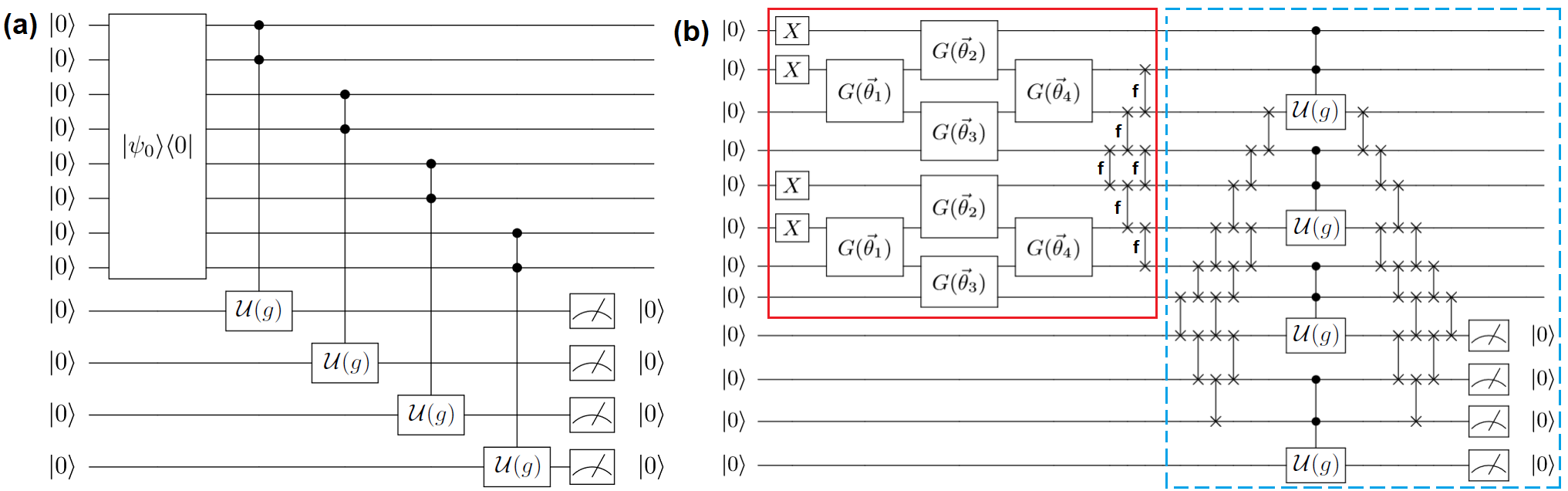}
\caption{(a) High-level scheme of routine to initialize GWF on a quantum computer, exemplified for a system of $N = 4$ sites. After the noninteracting ground state $\ket{\psi_0}$ is prepared, a single-qubit rotation is applied to each auxiliary qubit if the two qubits that encode the occupations of the corresponding site are in state $\ket{11}$. Measuring the auxiliary qubits and retrieving only the trials that yield $0000$ confirms the Gutzwiller projection was executed. (b) Detailed scheme of routine to initialize GWF assuming linear qubit connectivity. Solid red box corresponds to preparation of $\ket{\psi_0}$, which is first initialized in the diagonal basis and then transformed back to the original one via a Givens rotation decomposition, followed by the reordering of the qubits by site instead of spin, which involves fermionic SWAPs. Dashed blue box is the Gutzwiller projection: network of SWAPs places auxiliary qubits next to control-qubits. A detailed description can be found in the Supplemental Material.
\label{fig2}}
\end{figure*}

Henceforth, the initialization of the GWF on a digital quantum computer will be discussed. To the best of our knowledge, no routine has been proposed to accomplish this. The Jordan-Wigner mapping \cite{Jordan28} will be assumed, in which case the computational basis states $\ket{0}$ and $\ket{1}$ of each qubit encode the occupation of a spin orbital at a given site (unoccupied and occupied, respectively), yielding a total of $2N$ qubits to represent the wave function, where the factor of $2$ is due to the spin degeneracy. 

Unsurprisingly, the algorithm proceeds in two stages, as shown in Fig. \ref{fig2}: first, the noninteracting ground state $\ket{\psi_0}$ is prepared, then the Gutzwiller projection is applied. For the preparation of $\ket{\psi_0}$, we follow previous works\cite{Wecker15, Kivlichan18, Jiang18} based on the exploitation of the Thouless theorem\cite{Thouless60, Somma02} and the expression of the resulting unitary operation as a quantum circuit using a QR decomposition\cite{Horn85} via Givens rotations\cite{Press07}. As for the implementation of the Gutzwiller projector, the challenge associated with its non-unitary character needs to be overcome. This can be accomplished by embedding the projector in a larger unitary operation.

Since the Gutzwiller projection acts on each site separately, let us consider a single site, which is represented by two qubits, one for each spin. Let us add an auxiliary qubit, to which the single-qubit gate
\begin{equation}
    \mathcal{U}(g) = \begin{pmatrix}
     1-g & -\sqrt{2g - g^2}\\
     \sqrt{2g - g^2} & 1-g
    \end{pmatrix}
\end{equation}
is applied if and only if the two qubits that encode the occupations of the site are in state $\ket{11}$. Hence, for a given arbitrary two-qubit state $\ket{\Phi} = c_{00}\ket{00} + c_{01}\ket{01} + c_{10}\ket{10} + c_{11}\ket{11}$, the state after the $cc\, \mathcal{U}(g)$ reads:
\begin{align*}
cc\, \mathcal{U} (\ket{\Phi} \otimes \ket{0}) & = \left(c_{00}\ket{00}+c_{01}\ket{01}+c_{10}\ket{10}\right)\otimes \ket{0} \\
& + c_{11}\ket{11} \otimes \left((1-g)\ket{0} +\sqrt{2g-g^2}\ket{1}\right)
\end{align*}
Once the ancilla is measured in the computational basis, the state of the 
main register is collapsed into either
\begin{equation*}
\ket{\Phi_0} = c_{00}\ket{00} + c_{10}\ket{01} + c_{10}\ket{01} + c_{11} (1-g) \ket{11}
\end{equation*}
or
\begin{equation*}
\ket{\Phi_1} = c_{11} \sqrt{2g-g^2} \ket{11},
\end{equation*}
where the subscript denotes the outcome of the readout. Notice that $\ket{\Phi_0}$ coincides with the action of the Gutzwiller projection on the initial state $\ket{\Phi}$. Hence, to guarantee that the Gutzwiller projection is applied to this site, the auxiliary qubit must be initialized in $\ket{0}$ and measured in $\ket{0}$ after the application of $cc\, \mathcal{U}(g)$. The projection method is thus non-deterministic.

The application of the Gutzwiller projector to the whole wave function merely amounts to repeating this procedure for every site, as illustrated in Fig. 2(a). For the sake of clarity, let us compute its action on the following four-site wave function explicitly: 
\begin{align*}
    \ket{\psi} & = a\ket{\uparrow, \downarrow, \uparrow, \downarrow} +
                 b\ket{\uparrow, \uparrow, \downarrow, \downarrow} +
                 c\ket{\uparrow \downarrow, 0, \uparrow, \downarrow} \\
                 & + d\ket{\uparrow, \downarrow, \uparrow \downarrow, 0} +
                 e\ket{\uparrow \downarrow, 0, \uparrow \downarrow, 0} +
                 f\ket{\uparrow \downarrow, 0, 0, \uparrow \downarrow}.
\end{align*}
First, let us obtain the expected outcome by applying the Gutzwiller projector $\hat{P}_G(g) \equiv \prod_{i = 1}^{N} \mathbb{1} - g \hat{n}_{i\uparrow} \hat{n}_{i\downarrow}$ to $\ket{\psi}$:
\begin{align*}
    \hat{P}_G(g) \ket{\psi} = & \; a\ket{\uparrow, \downarrow, \uparrow, \downarrow} +
                              b\ket{\uparrow, \uparrow, \downarrow, \downarrow} \\
                              & + (1-g) \big[ c\ket{\uparrow \downarrow, 0, \uparrow, \downarrow}
                              + d\ket{\uparrow, \downarrow, \uparrow \downarrow, 0} \big] \\
                              & + (1-g)^2 \big[ e\ket{\uparrow \downarrow, 0, \uparrow \downarrow, 0}
                              + f\ket{\uparrow \downarrow, 0, 0, \uparrow \downarrow} \big].
\end{align*}
In words, the amplitude of the basis states with $n$ doubly-occupied sites is reduced by a factor of $(1-g)^n$. After normalization, the basis states with no doubly-occupied sites have a greater amplitude than originally. Let us now compare this to the action of the proposed quantum routine on $\ket{\psi}$. Adding the four ancillas initially in $\ket{0000}$, the wave function after the application of the four $cc\, \mathcal{U}$ but before the measurement of the ancillas is

\begin{widetext}
\begin{equation*}
\begin{aligned}
  & \quad \ket{0000} \otimes \Big( a\ket{\uparrow, \downarrow, \uparrow, \downarrow} + b\ket{\uparrow, \uparrow, \downarrow, \downarrow} + (1-g) c\ket{\uparrow \downarrow, 0, \uparrow, \downarrow} \\
  & \quad \quad \quad \quad \quad \; \; + (1-g) d\ket{\uparrow, \downarrow, \uparrow \downarrow, 0} + (1-g)^2 e\ket{\uparrow \downarrow, 0, \uparrow \downarrow, 0} + (1-g)^2 f\ket{\uparrow \downarrow, 0, 0, \uparrow \downarrow} \Big) \\ 
  & + \ket{0001} \otimes (1-g) \sqrt{2g -g^2} f\ket{\uparrow \downarrow, 0, 0, \uparrow \downarrow} + \ket{0010} \otimes \Big( \sqrt{2g -g^2} d\ket{\uparrow, \downarrow, \uparrow \downarrow, 0} + (1-g) \sqrt{2g -g^2} e\ket{\uparrow \downarrow, 0, \uparrow \downarrow, 0} \Big) \\
  & + \ket{1000} \otimes \Big( \sqrt{2g-g^2} c\ket{\uparrow \downarrow, 0, \uparrow, \downarrow} + (1-g) \sqrt{2g - g^2} e\ket{\uparrow \downarrow, 0, \uparrow \downarrow, 0} + (1-g) \sqrt{2g - g^2} f\ket{\uparrow \downarrow, 0, 0, \uparrow \downarrow} \Big) \\
  & + \ket{1001} \otimes (2g - g^2) f\ket{\uparrow \downarrow, 0, 0, \uparrow \downarrow} + \ket{1010} \otimes (2g - g^2) e\ket{\uparrow \downarrow, 0, \uparrow \downarrow, 0}.
\end{aligned}
\end{equation*}
\end{widetext}
The part of the wave function associated with the state $\ket{0000}$ in the auxiliary register coincides with $\hat{P}_G(g) \ket{\psi}$. Hence, all ancillas have to be measured in $\ket{0}$ to confirm the Gutzwiller projection was applied to the full state.

Given this method to apply the Gutzwiller projection on quantum hardware, the only piece left in the initialization of the GWF is the determination of the optimal value of $g$. This can be accomplished by minimizing the energy explicitly on the quantum computer, computing the expectation value of the Hamiltonian either via QPE or by decomposing it in the Pauli basis, as in VQE. Nevertheless, one can deduce the optimal value of $g$ for a large system by extrapolating from small system simulations carried out on a conventional computer. This is indeed the case for the one-dimensional FHM at half-filling, for which minor variations in the $g(U)$ relation are observed as the size of the chain varies. The same extrapolation should be possible for rectangular and square lattices.

In the remainder of this paper, we will discuss the scalability of the quantum routine herein proposed to initialize the GWF. In particular, the scaling of the relevant quantum circuit metrics (depth, width and number of CNOTs) and of the number of repetitions due to the non-deterministic nature of the method will be detailed.

Regarding the implementation of the quantum circuit for the Gutzwiller projection, the resulting overhead is found to be analogous to that associated with the preparation of the noninteracting ground state. Given $N$ lattice sites, $N$ additional qubits are required to perform the Gutzwiller projection, so the circuit width is $3N$ rather than $2N$. As for the circuit depth, although the controlled-rotations all act on independent trios of qubits, thereby allowing for their execution in parallel, the qubit connectivity must be taken into account, as the auxiliary qubits have to be placed in a position that is connected to the respective pair of control-qubits (cf. Fig. 2(b)). Making the realistic assumption of linear qubit connectivity\footnote{This assumption is particularly relevant for quantum computers based on superconducting circuits, for which the architectures typically only include linear connections for most qubits. As for trapped-ion quantum computers, there is, in principle, the possibility of achieving all-to-all connectivity, but, at the current state of development of the hardware, this comes at the cost of lower gate fidelities and greater execution times. Of course, if these technical limitations are overcome, trapped-ion quantum computers will allow to implement this Gutzwiller projection routine with negligible circuit depth overhead.}, the Gutzwiller projection requires a total of $\mathcal{O}(6N^2 + 18N)$ CNOTs and a circuit depth\footnote{The estimation of the circuit depth only includes CNOT gates, since two-qubit gates have considerably greater execution times and error rates than single-qubit gates.} of $\mathcal{O}(12N + 12)$, which compares to the $\mathcal{O}(4N^2 - 2N)$ CNOTs (at half-filling) and circuit depth of $\mathcal{O}(8N - 8)$ (at any filling) for the initialization of $\ket{\psi_0}$ (cf. Supplementary Material for a detailed discussion). It should be noted that, as shown in Fig. 2(b), in the initialization of $\ket{\psi_0}$ we include the reordering of the $2N$ qubits by site instead of spin: the noninteracting FHM is spin-polarized, so it is more practical to initialize $\ket{\psi_0}$ separately for each spin, but the propagator of the full Hamiltonian is more effectively implemented if the two qubits that represent the same site are next to each other\cite{Cai20, Cade20}. In summary, the circuit depth, width and number of CNOTs corresponding to the Gutzwiller projection are just a factor of $3/2$ greater than those for the preparation of $\ket{\psi_0}$.

\begin{table}    
\begin{tabular}{| l | p{1.5cm} | p{1.5cm} | p{1.5cm} | l |}
    \hline
     & $\textbf{N = 10}$ & $\textbf{N = 20}$ & $\textbf{N = 30}$ & $\textbf{N = 40}$ \\ \hline
    $\textbf{U/t = 1}$ & $2.7$ & $6.5$ & $16$ & $39$ \\ \hline
    $\textbf{U/t = 5}$ & $29$ & $940$ & $30,700$ & $1,000,000$ \\ \hline
    $\textbf{U/t = 10}$ & $63$ & $5,900$ & $550,000$ & $51,000,000$ \\ \hline
    $\textbf{U/t = 30}$ & $77$ & $9,000$ & $1,000,000$ & $120,000,000$ \\ \hline
    $\textbf{U/t = 50}$ & $78$ & $9,200$ & $1,100,000$ & $130,000,000$ \\ \hline
\end{tabular}
\caption{\label{Table1} Average number of repetitions required to prepare GWF on a digital quantum computer for one-dimensional FHM at half-filling with open boundary conditions. Values corresponding to $N = 10$ sites were obtained directly from simulation, while remaining ones were extrapolated from simulations of chains with $N = 2, 4, 6, 8, 10, 12$ sites. Cf. Supplemental Material for further details.}
\end{table}

\begin{table}  
\begin{tabular}{| p{1.5cm} | p{1.5cm} | p{1.5cm} |  p{1.5cm} | l |}
    \hline
    \textbf{N} & $\textbf{10}$ & $\textbf{20}$ & $\textbf{30}$ & $\textbf{40}$ \\ \hline
    $\mathbf{\ket{\psi_0}}$ & $11$ & $120$ & $1,500$ & $17,000$ \\ \hline
    $\mathbf{\ket{\psi_{MF}}}$ & $7.8$ & $70$ & $580$ & $5,000$ \\ \hline
    $\mathbf{\ket{\psi_{G}}^*}$ & $1.1$ & $1.2$ & $1.4$ & $1.6$ \\ \hline
    $\mathbf{\ket{\psi_{G}}^{**}}$ & $69$ & $7,100$ & $770,000$ & $82,000,000$ \\ \hline
\end{tabular}
\caption{\label{Table2} Estimate of average number of repetitions due to the choice of initial state $\ket{\psi_{initial}}$ that are required to find exact ground state $\ket{\psi_{exact}}$ of Fermi-Hubbard chain with $U/t = 10$ at half-filling via QPE or QITE. The initial states considered are the noninteracting ground state ($\ket{\psi_0}$), the self-consistent mean-field ground state ($\ket{\psi_{MF}}$) and the GWF, the latter under two assumptions: excluding repetition overhead to initialize it ($\ket{\psi_{G}}^{*}$), and including this overhead ($\ket{\psi_{G}}^{**}$). The number of trials due to the choice of initial state in QPE and QITE is estimated as $1/|\braket{\psi_{exact} | \psi_{initial}}|^2$ (cf. Supplemental Material for a detailed explanation). Values corresponding to $N = 10$ sites were obtained directly from simulation, while remaining ones were extrapolated from simulations of chains with $N = 2, 4, 6, 8, 10, 12$ sites.}
\end{table}

We now estimate the average number of times the quantum algorithm needs to be repeated for all $N$ ancillas to be found in the $\ket{0}$ state, thus ensuring a successful preparation of the GWF. This quantity depends both on $N$ and the value of $g$. In turn, $g$ depends on $U$. We have carried out this estimate numerically for chains with $N = 2, 4, 6, 8, 10, 12$ sites, and then extrapolated to larger chains. Table I shows the number of repetitions required to initialize the GWF for the FHM at half-filling for multiple chain sizes. Even for as many as $N = 40$ sites, the number of repetitions, though large, is clearly feasible for all $U/t$ values. We note that, in a similar spirit to VQE, this Gutzwiller projection scheme corresponds to a relatively shallow circuit that must be repeated multiple times. However, whereas in VQE these repetitions serve to explore the parameter space in search of the global minimum, in the Gutzwiller projection their cause is the non-deterministic nature of the routine.
 
However, if the GWF is needed as a starting point to carry out QPE or QITE, the improved fidelity does not offset the additional number of repetitions. This is illustrated in Table II, which compares the average number of repetitions due to the choice of initial state that are required to find the exact ground state, taking the initial state to be the noninteracting ground state ($\ket{\psi_0}$), the mean-field ground state ($\ket{\psi_{MF}}$) and the GWF, the latter case under two assumptions: excluding the repetition overhead to prepare the GWF ($\ket{\psi_{G}}^*$), and including this overhead ($\ket{\psi_{G}}^{**}$). The repetition overhead overwhelms the savings arising from replacing $\ket{\psi_0}$ or $\ket{\psi_{MF}}$ by $\ket{\psi_{G}}$. Likewise, the potential reduction of the number of layers or parameter updates in VQE arising from taking the GWF as the reference state will probably not compensate the increase in the number of repetitions, unless the barren plateau problem is especially acute.

One potentially promising route to tackle the repetition overhead issue is to combine this non-deterministic scheme with quantum amplitude amplification\cite{Brassard97, vanDyke21} for a suitable choice of oracle. This could lead to a trade-off between the number of repetitions and the circuit depth, thus making it a more viable solution for long-term quantum simulation algorithms such as QPE. Another line of research to be explored that could be especially relevant for near-term quantum simulation algorithms such as VQE is the preparation of the GWF via an appropriate parameterized quantum circuit.

In conclusion, the GWF was proposed as a promising trial state for the digital quantum simulation of strongly-correlated electrons. We developed the first routine to prepare the GWF on a gate-based quantum computer, the circuit depth and width requirements of which are similar to those associated with the initialization of the noninteracting ground state. The non-deterministic nature of this scheme leads to a repetition overhead that, though viable for the implementation of the GWF on its own, must be reduced in order to integrate it in general quantum simulation algorithms.

{\it Acknowledgements}. B.M. acknowledges support from the FCT PhD scholarship no. SFRH/BD/08444/2020. JFR acknowledges financial support from the Spanish Government (grant no. PID2019-109539GB-C41), and Generalitat Valenciana funding Prometeo 2017/139. The authors acknowledge use of the QuSpin\cite{QuSpin} package for the exact diagonalization calculations and of Qiskit\cite{Qiskit} for the in-silico simulations of the quantum circuits.

\pagebreak

\onecolumngrid


\begin{center}
  \textbf{\large Supplemental Material for\\ ``Gutzwiller wave function on a digital quantum computer''}\\[.2cm]
\end{center}

\setcounter{equation}{0}
\setcounter{figure}{0}
\setcounter{table}{0}
\renewcommand{\theequation}{S\arabic{equation}}
\renewcommand{\thefigure}{S\arabic{figure}}

\section{Basis Gate Decomposition and Scaling of Quantum Circuit Depth and Width}

This section details how to decompose each operation in the quantum circuit shown in Fig. 2 of the main text in terms of basis gates. The reference two-qubit gate assumed is the CNOT \cite{NielsenChuang}. The initialization of the Gutzwiller Wave Function (GWF) on a digital quantum computer comprises two parts: the preparation of the noninteracting ground state $\ket{\psi_0}$ and the Gutzwiller projection $\hat{P}_G(g)$.

\vspace{0.3cm}
\centerline{\textit{Preparation of Noninteracting Ground State}}
\vspace{0.3cm}

If the noninteracting Hamiltonian is spin-polarized (which is generally the case, since hopping is not normally assumed to mix different spin states), the preparation of $\ket{\psi_0}$ can be treated separately for the two spins $\uparrow$ and $\downarrow$, which effectively leads to two subcircuits that can be executed in parallel, as shown in Fig. 2(b) of the main text for the spin-balanced case $N_{\uparrow} = N_{\downarrow} = 2$ (in which case the two subcircuits are equal). We can thus consider only one spin state in the following discussion. 

\begin{figure}[h]
\centering{\
\includegraphics[width=1.0\linewidth]{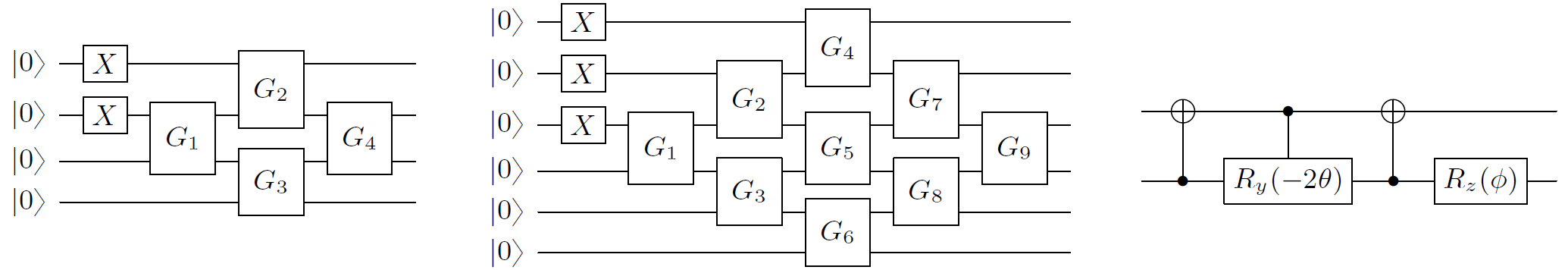}
}
\caption{Scheme of quantum circuits to initialize spin-polarized noninteracting ground state of a given Hamiltonian for $N = 4$ (left) and $N = 6$ (center) sites. In both examples, $N_s = N/2$, where $s = \uparrow, \downarrow$ is the selected spin. For a different number of electrons, the number of NOT gates in the initialization of the reference Slater determinant and the number of Givens rotations $G_i$ in the decomposition of the unitary transformation between the two bases would change. The two-qubit operation $G_i$ can be decomposed in terms of more elementary gates as shown on the right-hand side.}
\label{Fig1}
\end{figure}

Fig. \ref{Fig1} above shows a scheme of the quantum circuits to initialize the noninteracting ground state $\ket{\psi_0}$ for $N_s = N/2$, with $N = 4$ (left) and $N = 6$ (center) sites. The occupations of the single-particle orbitals are encoded in the qubits from top to bottom in increasing order of energy. Hence, the Slater determinants corresponding to $\ket{\psi_0}$ in the diagonal basis are $\ket{1100}$ and $\ket{111000}$, respectively. These two reference states are initialized via the application of the NOT gates at the start of the circuits. The remainder of the circuits amounts to the unitary transformation between the diagonal basis and the original one, and is decomposed in terms of Givens rotations $G_i$. The basis gate decomposition of each $G_i$ is shown on the right-hand-side of Fig. \ref{Fig1}; the only operation in this quantum circuit that is neither a CNOT nor a single-qubit gate is the controlled-$R_y(-2\theta)$, which can be decomposed as

\begin{figure}[h]
\centering{\
\Qcircuit @C=1em @R=1.2em {
  \lstick{} & \qw & \ctrl{1} & \qw & \ctrl{1} & \qw \\
  \lstick{} & \gate{R_y(-\theta)} & \targ & \gate{R_y(\theta)} & \targ & \qw 
} \par}
\label{Givens_circ}
\end{figure}

\noindent Hence, each $G_i$ operation requires $4$ CNOTs. Since a total of $N_s (N-N_s)$ $G_i$ operations are required to perform the basis transformation, this gives a total of $2 \cdot N_s (N - N_s) \cdot 4 = 8 N_s (N-N_s)$ CNOTs, where the extra factor of $2$ is due to the spin degeneracy, assuming, as usual, that $N_{\uparrow} = N_{\downarrow}$. At half-filling, $N_s = N/2$, so the total number of CNOTs in this case is $2N^2$. As for the circuit depth, the unitary transformation that relates the two bases involves $N-1$ layers of $G_i$ operations regardless of the filling level. Ignoring single-qubit gates, which have execution times and error rates far lower than those of CNOTs, and noting that each $G_i$ involves $4$ CNOTs that must be applied sequentially, the total circuit depth for this part is $4(N-1)$. Note that there is no factor of $2$ due to the spin degeneracy, since the two subcircuits, one for each spin, can be executed in parallel.

\begin{figure}
\centering{\
\includegraphics[width=0.5\linewidth]{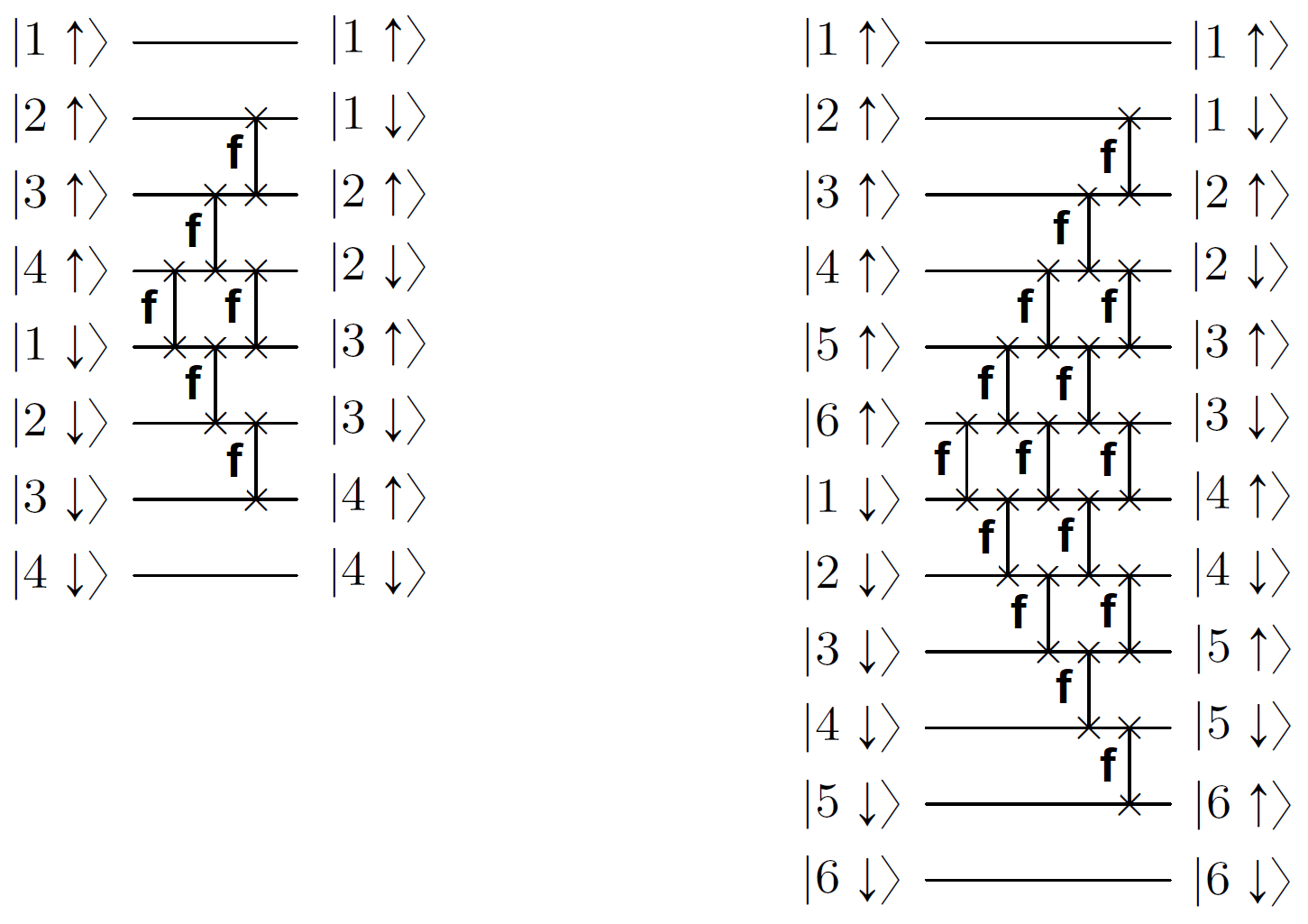}
\par}
\caption{Scheme of quantum circuit to reorder qubits by site instead of spin after the initialization of the spin-polarized noninteracting ground states in parallel for $N = 4, 6$ sites. After the application of this array of fermionic-SWAPs, the two qubits that encode the occupations of the two spin-orbitals associated with the same site are next to each other.}
\label{Fig2}
\end{figure}

Before the noninteracting ground state $\ket{\psi_0}$ is used as the starting point of a quantum simulation algorithm such as Quantum Phase Estimation (QPE) \cite{Kitaev95} or the Variational Quantum Eigensolver (VQE) \cite{Peruzzo14}, it is common practice to reorder the qubits by site instead of spin (i.e. place the two qubit that encode the occupations of the two spin states at the same site next to each other) when dealing with lattice models of correlated electrons with on-site Hubbard interactions \cite{Cade20, Cai20}. This assumes, of course, that the qubit connectivity is limited, namely that qubits are only connected linearly. This qubit reordering therefore allows to reduce the circuit depth of the implementation of the propagator $e^{-i \mathcal{H} t/\hbar}$, and it can be accomplished via the quantum circuits shown in Fig. \ref{Fig2} for $N = 4, 6$ sites. To account for the exchange anti-symmetry, fermionic-SWAPs are required, which are analogous to conventional SWAPs, except for the extra minus sign applied to the basis state $\ket{11}$. The basis gate decomposition of a fermionic-SWAP is
\begin{equation*}
    \Qcircuit @C=1em @R=1.2em {
  \lstick{} & \ctrl{1} & \targ & \ctrl{1} & \gate{H} & \targ & \gate{H} & \qw \\
  \lstick{} & \targ & \ctrl{-1} & \targ & \qw & \ctrl{-1} & \qw & \qw 
    }
\end{equation*}
Hence, the reordering of the $2N$ qubits by site involves a total of $N(N-1)/2$ fermionic-SWAPs, each taking $4$ CNOTs, which brings the total CNOT count to $2N(N-1)$. As for the circuit depth, $N-1$ layers of fermionic-SWAPs are needed and each takes $4$ CNOTs, so the total circuit depth (in CNOTs) is of $4(N-1)$. 

To sum up, the initialization of the noninteracting ground state corresponds to a total circuit depth of $8N - 8$ CNOTs and a total number of CNOTs of $8N_s(N-N_s) + 2N(N-1)$ at any filling, or $4N^2 - 2N$ at half-filling.

\vspace{0.3cm}
\centerline{\textit{Gutzwiller Projection}}
\vspace{0.3cm}

Ignoring any qubit connectivity constraints for the moment, all $N$ controlled-controlled-$\mathcal{U}(g)$ corresponding to the Gutzwiller projection can be implemented in parallel, since each acts on its own set of three qubits. Since $\mathcal{U}(g) = R_y(\alpha(g))$, with $\alpha(g) = 2 \arctan (\sqrt{2g - g^2} / (1-g))$, the basis gate decomposition of the controlled-controlled-$\mathcal{U}(g)$ can be derived immediately from that of the controlled-$R_y(\theta)$ (cf. previous page):

\begin{figure}[h]
\centering{\
\Qcircuit @C=1em @R=1.2em {
  \lstick{} & \qw & \ctrl{1} & \qw & \ctrl{1} & \qw \\
  \lstick{} & \qw & \ctrl{1} & \qw & \ctrl{1} & \qw \\
  \lstick{} & \gate{R_y(\alpha(g)/2)} & \targ & \gate{R_y(-\alpha(g)/2)} & \targ & \qw 
} \par}
\end{figure}

\noindent Hence, each controlled-controlled-$\mathcal{U}(g)$ requires two Toffoli gates \cite{Toffoli80}, each of which can be implemented as \cite{NielsenChuang}:

\begin{figure}[h]
\centering{\
\Qcircuit @C=1em @R=1.2em {
  \lstick{} & \qw & \qw & \qw & \ctrl{2} & \qw & \qw & \qw & \ctrl{2} & \qw & \ctrl{1} & \gate{T} & \ctrl{1} & \qw \\
  \lstick{} & \qw & \ctrl{1} & \qw & \qw & \qw & \ctrl{1} & \qw & \qw & \gate{T} & \targ & \gate{T^{\dagger}} & \targ & \qw \\
  \lstick{} & \gate{H} & \targ & \gate{T^{\dagger}} & \targ & \gate{T} & \targ & \gate{T^{\dagger}} & \targ & \gate{T} & \gate{H} & \qw & \qw & \qw
} \par}
\end{figure}

\noindent In summary, assuming all-to-all connectivity, the Gutzwiller projection corresponds to a circuit depth of $2 \cdot 6 = 12$ CNOTs (i.e. independent of the size of the system) and a total number of CNOTs of $12N$. 

However, in current quantum processors, the assumption of linear qubit connectivity is far more realistic than all-to-all connectivity. In such case, there are two main changes relative to the analysis above: first, the auxiliary qubits must be moved next to the two qubits associated with the respective site; second, in the basis gate decomposition of the Toffoli gate shown above, the middle and bottom qubits must be swapped in order to execute the first four CNOTs (assuming only the top/middle and middle/bottom pairs of qubits are connected to each other). The latter change, shown in the scheme below, leads to a total of $2 \cdot 3 + 6 = 12$ CNOTs, as opposed to the original $6$.

\begin{figure}[h]
\centering{\
\Qcircuit @C=1em @R=1.2em {
  \lstick{} & \qw & \qw & \qw & \qw & \qw & \qw & \ctrl{1} & \qw & \qw & \qw & \ctrl{1} & \qw & \qw & \qw & \qw & \ctrl{1} & \gate{T} & \ctrl{1} & \qw \\
  \lstick{} & \ctrl{1} & \targ & \ctrl{1} & \gate{H} & \targ & \gate{T^{\dagger}} & \targ & \gate{T} & \targ & \gate{T^{\dagger}} & \targ & \ctrl{1} & \targ & \ctrl{1} & \gate{T} & \targ & \gate{T^{\dagger}} & \targ & \qw \\
  \lstick{} & \targ & \ctrl{-1} & \targ & \qw & \ctrl{-1} & \qw & \qw & \qw & \ctrl{-1} & \qw & \qw & \targ & \ctrl{-1} & \targ & \gate{T} & \gate{H} & \qw & \qw & \qw
} \par}
\end{figure}

\noindent Hence, with linear qubit connectivity, the implementation of the $N$ controlled-controlled-$\mathcal{U}(g)$ gates involves $24 N$ CNOTs in total and a circuit depth of $24$ CNOTs.

\begin{figure}[h]
\centering{\
\Qcircuit @C=0.8em @R=1.5em {
\lstick{|1\uparrow\rangle} & \qw & \qw & \qw & \qw & \qw & \qw & \qw & \hspace{0.5cm} |1\uparrow\rangle\\
\lstick{|1\downarrow\rangle} & \qw & \qw & \qw & \qw & \qw & \qw & \qw & \hspace{0.5cm} |1\downarrow\rangle\\
\lstick{|2\uparrow\rangle} & \qw & \qw & \qw & \qw & \qw & \qswap & \qw & \hspace{0.5cm} |A1\rangle\\
\lstick{|2\downarrow\rangle} & \qw & \qw & \qw & \qw & \qswap & \qswap \qwx & \qw & \hspace{0.5cm} |2\uparrow\rangle\\
\lstick{|3\uparrow\rangle} & \qw & \qw & \qw & \qswap & \qswap \qwx & \qw & \qw & \hspace{0.5cm} |2\downarrow\rangle\\
\lstick{|3\downarrow\rangle} & \qw & \qw & \qswap & \qswap \qwx & \qswap & \qw & \qw & \hspace{0.5cm} |A2\rangle\\
\lstick{|4\uparrow\rangle} & \qw & \qswap & \qswap \qwx & \qswap & \qswap \qwx & \qw & \qw & \hspace{0.5cm} |3\uparrow\rangle\\
\lstick{|4\downarrow\rangle} & \qswap & \qswap \qwx & \qswap  & \qswap \qwx & \qw & \qw & \qw & \hspace{0.5cm} |3\downarrow\rangle\\
\lstick{|A1\rangle} & \qswap \qwx & \qswap & \qswap \qwx & \qswap & \qw & \qw & \qw & \hspace{0.5cm} |A3\rangle\\
\lstick{|A2\rangle} & \qw & \qswap \qwx & \qswap & \qswap \qwx & \qw & \qw & \qw & \hspace{0.5cm} |4\uparrow\rangle\\
\lstick{|A3\rangle} & \qw & \qw & \qswap \qwx & \qw & \qw & \qw & \qw & \hspace{0.5cm} |4\downarrow\rangle \\
\lstick{|A4\rangle} & \qw & \qw & \qw & \qw & \qw & \qw & \qw & \hspace{0.5cm} |A4\rangle \\
}
\hspace{3.5cm}
\Qcircuit @C=0.8em @R=1.5em {
\lstick{|1\uparrow\rangle} & \qw & \qw & \qw & \qw & \qw & \qw & \qw & \qw & \qw & \qw & \qw & \hspace{0.5cm} |1\uparrow\rangle\\
\lstick{|1\downarrow\rangle} & \qw & \qw & \qw & \qw & \qw & \qw & \qw & \qw & \qw & \qw & \qw & \hspace{0.5cm} |1\downarrow\rangle\\
\lstick{|2\uparrow\rangle} & \qw & \qw & \qw & \qw & \qw & \qw & \qw & \qw & \qw & \qswap & \qw & \hspace{0.5cm} |A1\rangle\\
\lstick{|2\downarrow\rangle} & \qw & \qw & \qw & \qw & \qw & \qw & \qw & \qw & \qswap & \qswap \qwx & \qw & \hspace{0.5cm} |2\uparrow\rangle\\
\lstick{|3\uparrow\rangle} & \qw & \qw & \qw & \qw & \qw & \qw & \qw & \qswap & \qswap \qwx & \qw & \qw & \hspace{0.5cm} |2\downarrow\rangle\\
\lstick{|3\downarrow\rangle} & \qw & \qw & \qw & \qw & \qw & \qw & \qswap & \qswap \qwx & \qswap & \qw & \qw & \hspace{0.5cm} |A2\rangle\\
\lstick{|4\uparrow\rangle} & \qw & \qw & \qw & \qw & \qw & \qswap & \qswap \qwx & \qswap & \qswap \qwx & \qw & \qw & \hspace{0.5cm} |3\uparrow \rangle\\
\lstick{|4\downarrow\rangle} & \qw & \qw & \qw & \qw & \qswap & \qswap \qwx & \qswap & \qswap \qwx & \qw & \qw & \qw & \hspace{0.5cm} |3\downarrow\rangle\\
\lstick{|5\uparrow\rangle} & \qw & \qw & \qw & \qswap & \qswap \qwx & \qswap & \qswap \qwx & \qswap & \qw & \qw & \qw & \hspace{0.5cm} |A3\rangle\\
\lstick{|5\downarrow\rangle} & \qw & \qw & \qswap & \qswap \qwx & \qswap & \qswap \qwx & \qswap & \qswap \qwx & \qw & \qw & \qw & \hspace{0.5cm} |4\uparrow\rangle\\
\lstick{|6\uparrow\rangle} & \qw & \qswap & \qswap \qwx & \qswap & \qswap \qwx & \qswap & \qswap \qwx & \qw & \qw & \qw & \qw & \hspace{0.5cm} |4\downarrow\rangle\\
\lstick{|6\downarrow\rangle} & \qswap & \qswap \qwx & \qswap & \qswap \qwx & \qswap & \qswap \qwx & \qswap & \qw & \qw & \qw & \qw & \hspace{0.5cm} |A4\rangle\\
\lstick{|A1\rangle} & \qswap \qwx & \qswap & \qswap \qwx & \qswap & \qswap \qwx & \qswap & \qswap \qwx & \qw & \qw & \qw & \qw & \hspace{0.5cm} |5\uparrow\rangle\\
\lstick{|A2\rangle} & \qw & \qswap \qwx & \qswap & \qswap \qwx & \qswap & \qswap \qwx & \qw & \qw & \qw & \qw & \qw & \hspace{0.5cm} |5\downarrow\rangle\\
\lstick{|A3\rangle} & \qw & \qw & \qswap \qwx & \qswap & \qswap \qwx & \qswap & \qw & \qw & \qw & \qw & \qw & \hspace{0.5cm} |A5\rangle \\
\lstick{|A4\rangle} & \qw & \qw & \qw & \qswap \qwx & \qswap & \qswap \qwx & \qw & \qw & \qw & \qw & \qw & \hspace{0.5cm} |6\uparrow\rangle \\
\lstick{|A5\rangle} & \qw & \qw & \qw & \qw & \qswap \qwx & \qw & \qw & \qw & \qw & \qw & \qw & \hspace{0.5cm} |6\downarrow\rangle \\
\lstick{|A6\rangle} & \qw & \qw & \qw & \qw & \qw & \qw & \qw & \qw & \qw & \qw & \qw & \hspace{0.5cm} |A6\rangle \\
}
\par}
\caption{Scheme of quantum circuit to place auxiliary qubits next to respective control-qubits to implement controlled-controlled-$\mathcal{U}(g)$ corresponding to Gutzwiller projection for $N = 4, 6$ sites assuming linear qubit connectivity constraint. The inverse of this network of SWAPs must be applied after the execution of the controlled-controlled-$\mathcal{U}(g)$ to return the $2N$ qubits in the main register to their ideal positions for the remainder of the quantum simulation.}
\label{Fig3}
\end{figure}

The placement of the auxiliary qubits next to the two respective control-qubits requires a network of conventional SWAPs such as the ones shown in Fig. \ref{Fig3} for $N = 4, 6$ sites. Note, however, that this network of SWAPs must be repeated after the execution of the controlled-controlled-$\mathcal{U}(g)$ gates so that the auxiliary qubits are returned to their original positions and, more importantly, the $2N$ qubits of the main register are ready for the remaining part of the quantum simulation algorithm to be implemented. In total, this therefore requires $2N(N-1)$ SWAPs, or $6N(N-1)$ CNOTs, and a circuit depth of $4(N-1)$ SWAPs, or $12(N-1)$ CNOTs. To sum up, the Gutzwiller projection requires a total of $6N^2 + 18N$ CNOTs and a circuit depth of $12N + 12$ CNOTs.

\vspace{0.3cm}
\centerline{\textit{Generalization to Arbitrary Geometries and Dimensions}}
\vspace{0.3cm}

It should be stressed that this analysis of the scaling of the circuit depth, width and total number of CNOTs of the quantum circuit corresponding to the Gutzwiller projection routine is agnostic to the geometry of the lattice considered. Regarding the preparation of the noninteracting ground state, what determines the number of Givens rotations required is simply the number of electrons and the number of single-particle orbitals, the latter being set by the number of lattice sites only. As for the Gutzwiller projection itself, being a local operator, it is clearly independent of the geometry. Finally, the SWAP networks exemplified schematically in Figs. \ref{Fig2} and \ref{Fig3} are applicable to lattices of arbitrary geometry and dimension. The only underlying assumption is that, at the moment of the preparation of the spin-polarized noninteracting states, the qubits corresponding to a given spin $s = \uparrow, \downarrow$ are already ordered according to the convention chosen for the labelling of the lattice sites, whatever that is from a spatial point of view.

\section{Scaling of Repetition Overhead}

The Gutzwiller projection routine herein proposed is probabilistic, in that the GWF is only initialized on the quantum computer if all auxiliary qubits are measured in $\ket{0}$. In this section, the average number of repetitions required to prepare the GWF is determined. The application of the $N$ controlled-controlled-$\mathcal{U}(g)$ gates on $\ket{0}^{\otimes N} \otimes \ket{\psi_0}$ gives $\ket{\psi'} \equiv \ket{0}^{\otimes N} \otimes \ket{\psi_G} + ...$, where $\ket{\psi_G} = \hat{P}_G(g) \ket{\psi_0}$ is the GWF and the ellipsis refers to the remaining terms of the wave function for which the auxiliary qubits are in a computational basis state other than $\ket{0}^{\otimes N}$. The probability that all auxiliary qubits are measured in the $\ket{0}$ state is given by
\begin{equation}
    P(N, g) = \sum_{\{ q_1, q_2, ..., q_{2N}\} \in \{ 0, 1\}^{\otimes 2N}} |\big(\bra{0}^{\otimes N} \otimes \bra{q_1 q_2 ... q_{2N}}\big) \ket{\psi'}|^2 = |\braket{\psi_G | \psi_G}|^2,
\end{equation}
where $\ket{q_1 q_2 ... q_{2N}}$ is the main $2N$-qubit register where $\ket{\psi_0}$ is initialized and $\ket{\psi_G}$ is later obtained. Notice that, due to the non-unitarity of $\hat{P}_G(g)$, $\ket{\psi_G}$ is not normalized, which is why the probability of preparing it is not generally equal to 1. The noninteracting ground state $\ket{\psi_0}$ at half-filling for a lattice of $N$ sites can be written as
\begin{equation}
    \ket{\psi_0} = \sum_{n = 0}^{N/2} \sum_{i = 1}^{M_n} c_i^{(n)} \ket{\phi_i^{(n)}},
\end{equation}
where $n$ is the number of doubly-occupied sites, and therefore $\ket{\phi_i^{(n)}}$ is a basis state with $n$ doubly-occupied sites, with $M_n = {N \choose n} {N-n \choose n} {N - 2n \choose (N-2n)/2}$ the number of basis states with $n$ doubly-occupied sites. The GWF is given by
\begin{equation}
    \ket{\psi_{G}} = \hat{P}_G(g) \ket{\psi_0} = \sum_{n = 0}^{N/2} \sum_{i = 1}^{N_n} (1-g)^n c_i^{(n)} \ket{\phi_i^{(n)}},
\end{equation}
in which case the probability of measuring the auxiliary qubits in $\ket{0}^{\otimes N}$ is
\begin{equation}
    P(N, g) = \sum_{n = 0}^{N/2} \sum_{i = 1}^{N_n} (1-g)^{2n} |c_i^{(n)}|^2.
\end{equation}
The average number of repetitions for a given number of lattice sites $N$ and Gutzwiller parameter $g$ is simply $1/P(N, g)$. As a sanity check, if we set $g = 0$, $\ket{\psi_G}$ is just $\ket{\psi_0}$, so $P = 1$ and the number of repetitions is $1$, as expected. 

\begin{figure}[h]
    \centering
    \includegraphics[width=0.85\linewidth]{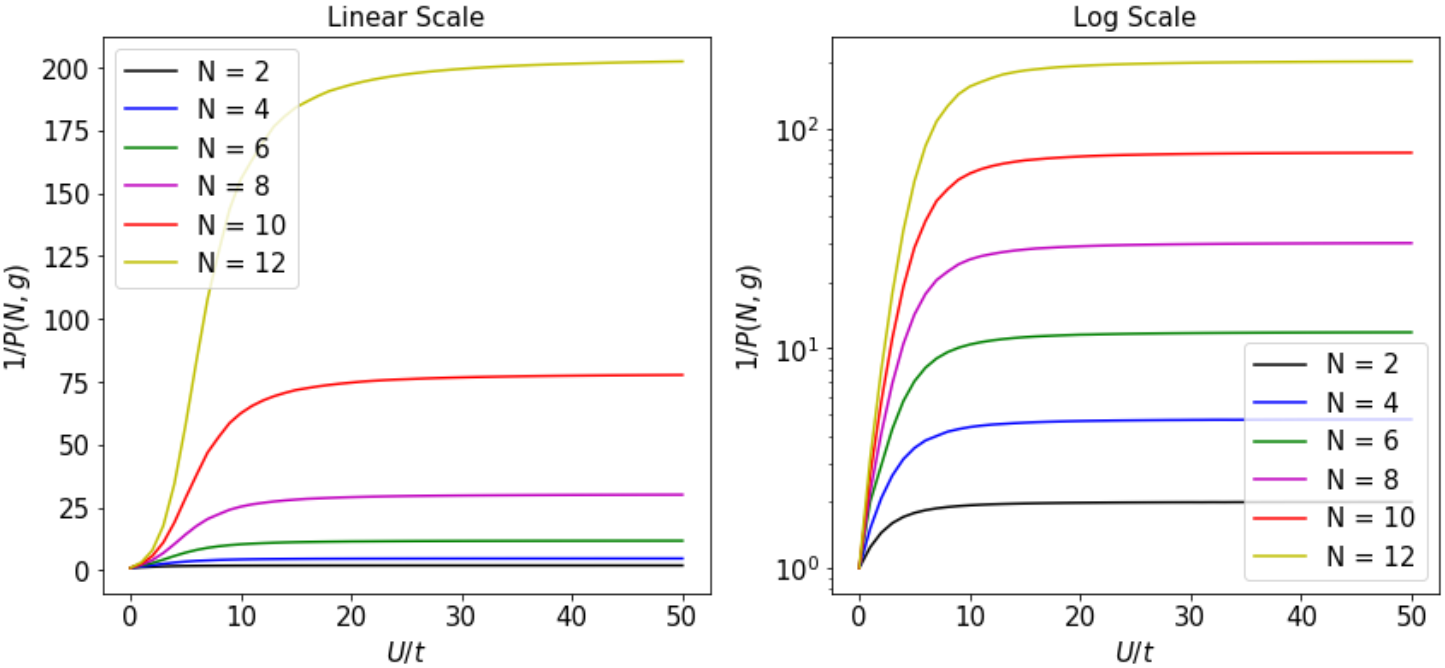}
    \caption{Average number of repetitions required to initialize GWF on quantum computer against normalized Hubbard parameter $U/t$ for the Fermi-Hubbard model in chains of varying size.}
    \label{Fig4}
\end{figure}

Fig. \ref{Fig4} shows the average number of repetitions required to prepare the GWF on quantum hardware against the normalized Hubbard parameter $U/t$ for the Fermi-Hubbard model in chains of varying size. In all cases, the optimal Gutzwiller parameter $g$ that minimizes the energy is considered. As is clear from the right-hand-side plot in logarithmic scale, the number of repetitions grows exponentially with the system size.

\begin{figure}[h]
    \centering
    \includegraphics[width=0.85\linewidth]{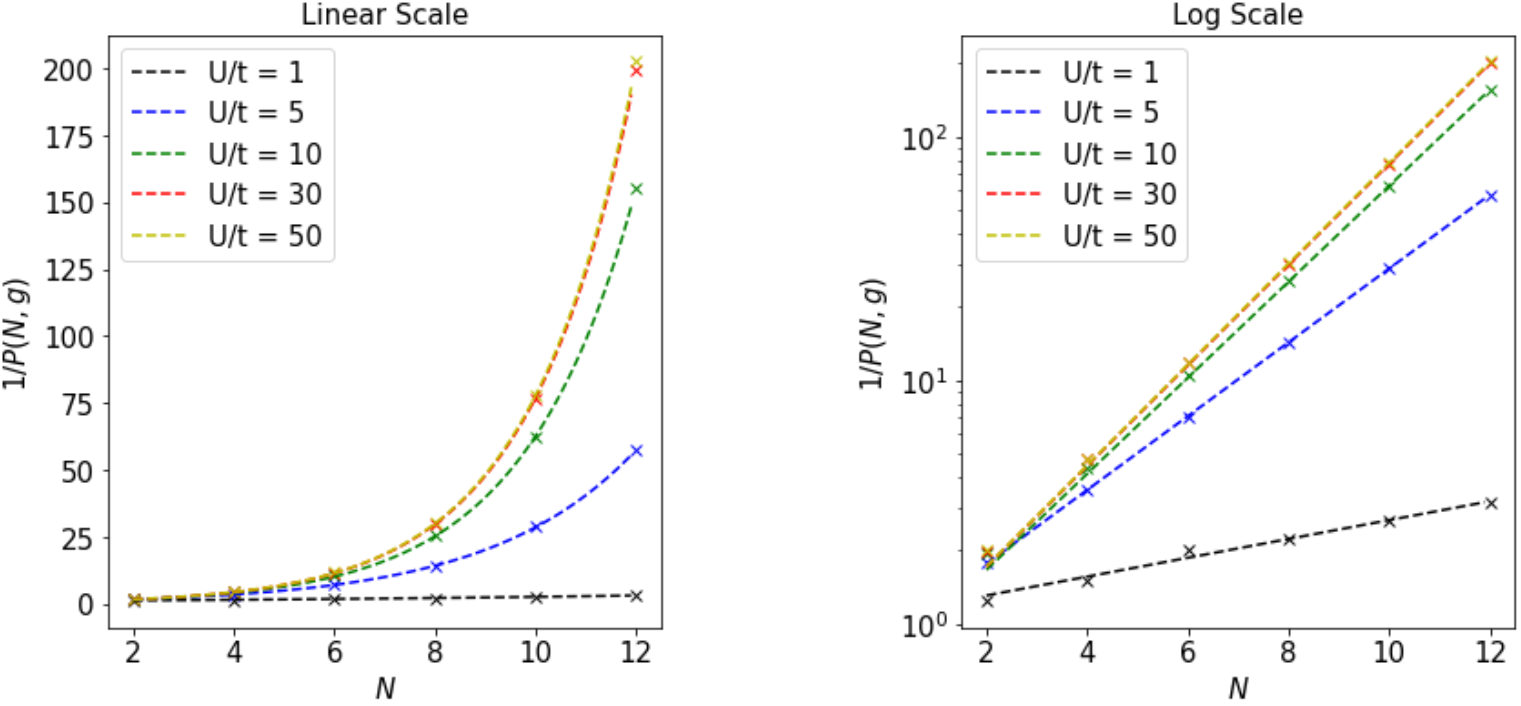}
    \caption{Extrapolation of average number of repetitions required to initialize the GWF on quantum hardware from results of simulations for Fermi-Hubbard model in chains of $N = 2, 4, 6, 8, 10, 12$ sites for different values of $U/t$. Numerical results were fitted to exponential function $c_1e^{c_2N}$, with free parameters $c_1$ and $c_2$.}
    \label{Fig5}
\end{figure}

In order to estimate the number of repetitions required for larger lattices where quantum advantage may be attained (cf. Table I in main text), an extrapolation of the results obtained from simulations with chains of $N = 2, 4, 6, 8, 10, 12$ sites was carried out for $U/t = 1, 5, 10, 30, 50$, as shown in Fig. \ref{Fig5}. 

\section{Scaling of Repetitions as a Function of Trial State for QPE and QITE}

The replacement of the noninteracting or self-consistent mean-field ground states by the GWF as the reference state for Quantum Phase Estimation (QPE) \cite{Kitaev95} or Quantum Imaginary Time Evolution (QITE) \cite{Motta20, McArdle191} could lead to significant savings of repetitions. In order to make an estimate of such savings for sufficiently large lattices where quantum advantage may be attained, an extrapolation analysis of the fidelity with respect to the exact ground state for the three reference states considered was carried out, from which the average number of repetitions due to the choice of initial state that are required to obtain the exact ground state was estimated as $1/|\braket{\psi_{trial} | \psi_{exact}}|^2$:

\begin{itemize}
\item{QPE takes the state $\ket{0}^{\otimes t} \otimes \ket{\psi_{exact}^{(n)}}$ to $\ket{\tilde{\epsilon}_n} \otimes \ket{\psi_{exact}^{(n)}}$, where the $t$-qubit register corresponds to the auxiliary qubits that encode the phase, $\ket{\psi_{exact}^{(n)}}$ is the $n^{\textrm{th}}$ exact eigenstate of the Hamiltonian $\mathcal{H}$, $\epsilon_n$ is the respective eigenvalue and $\tilde{\epsilon_n}$ is the estimate of $\epsilon_n$ accurate to $k$ bits of precision with probability of success $1 - s$, provided that the number of auxiliary qubits is chosen to be $t \geq k + \log (2 + 1/(2s))$ \cite{NielsenChuang}. It then follows by linearity that, if instead the input happens to be a linear combination of the eigenstates of the Hamiltonian, as in $\ket{0}^{\otimes t} \otimes \sum_n \braket{\psi_{exact}^{(n)} | \psi_{trial} } \ket{\psi_{exact}^{(n)}}$, the probability of measuring $\epsilon_n$ accurate to $k$ bits and thus collapsing the main register onto the corresponding eigenstate is at least $|\braket{\psi_{exact}^{(n)} | \psi_{trial} }|^2 (1-s)$ \cite{NielsenChuang}.}
\item{In QITE, the trial state can also be expanded in the eigenbasis of the Hamiltonian, in which case the action of the (normalized) imaginary-time propagator is $\mathcal{N}(\tau) e^{-\epsilon_0 \tau} [ \braket{\psi_{exact} | \psi_{trial}} \ket{\psi_{exact}} + e^{-(\epsilon_1 - \epsilon_0) \tau} \braket{\psi_{exact}^{(1)} | \psi_{trial}} \ket{\psi_{exact}^{(1)}} + e^{-(\epsilon_2 - \epsilon_0) \tau} \braket{\psi_{exact}^{(2)} | \psi_{trial}} \ket{\psi_{exact}^{(2)}} + ... ]$, where $\mathcal{N}(\tau) = 1/\sqrt{\braket{\psi_{trial} | e^{-2\mathcal{H} \tau} | \psi_{trial}}}$ is the normalization factor. The probability of collapsing onto $\ket{\psi_{exact}}$ after imaginary time $\tau$ is $P(\tau) = \mathcal{N}^{2}(\tau) e^{-2\epsilon_0 \tau} |\braket{\psi_{exact} | \psi_{trial}}|^2$.}
\end{itemize}

\bibliographystyle{apsrev4-1}
\bibliography{biblio}

\end{document}